\documentclass{aastex63}

\usepackage{graphicx}
\graphicspath{ {./Images/} }
\usepackage{amsmath}	
\usepackage{amssymb}
\newcolumntype{P}[1]{>{\centering\arraybackslash}p{#1}}

\submitjournal{APJ}

\shorttitle{The Bimodal Distribution in Exoplanet Radii}
\shortauthors{Modirrousta-Galian et al.}

\begin{document}
	
	\title{The Bimodal Distribution in Exoplanet Radii: Considering Varying Core Compositions and $\rm H_{2}$ Envelope's Sizes}

	\correspondingauthor{Darius Modirrousta-Galian}
	\email{darius.modirrousta@inaf.it}
	
	\author[0000-0001-6425-9415]{Darius Modirrousta-Galian}
	\affiliation{INAF – Osservatorio Astronomico di Palermo, Piazza del Parlamento 1, I-90134 Palermo, Italy}
	\affiliation{University of Palermo, Department of Physics and Chemistry, Via Archirafi 36, Palermo, Italy}
	
	\author{Daniele Locci}
	\affiliation{INAF – Osservatorio Astronomico di Palermo, Piazza del Parlamento 1, I-90134 Palermo, Italy}
	
	\author{Giuseppina Micela}
	\affiliation{INAF – Osservatorio Astronomico di Palermo, Piazza del Parlamento 1, I-90134 Palermo, Italy}

	\begin{abstract}
		Several models have been introduced in order to explain the radius distribution in exoplanet radii observed by \citet{Fulton2017} with one peak at $\rm \sim 1.3 R_{\oplus} $ the other at $\rm \sim 2.4 R_{\oplus} $ and the minimum at $\rm \sim 1.75R_{\oplus} $. In this paper we focus on the hypothesis that the exoplanet size distribution is caused by stellar XUV-induced atmospheric loss. We evolve $10^{6}$ synthetic exoplanets by exposing them to XUV irradiation from synthetic ZAMS stars. For each planet we set a different interior composition which ranged from $\rm 100 \: wt\%$ Fe (very dense) through $\rm 100 \: wt\%$ $\rm MgSiO_{3}$ (average density) and to $\rm 100 \: wt\%$ $\rm H_{2}O$ ice (low density) with varying hydrogen envelop sizes which varied from $\rm 0 \: wt\%$ (a negligible envelop) to $\rm 100 \: wt\%$ (a negligible core). Our simulations were able to replicate the bimodal distribution in exoplanet radii. We argue that in order to reproduce the distribution by \citet{Fulton2017} it is mandatory for there to be a paucity of exoplanets with masses above $\rm \sim 8M_{\oplus}$. Furthermore, our best-fit result predicts an initial flat distribution in exoplanet occurrence for $\rm M_{P} \lesssim 8M_{\oplus}$ with a strong deficiency for planets with $\rm \lesssim 3M_{\oplus}$. Our results are consistent with the $\rm \sim 1.3R_{\oplus}$ radius peak mostly encompassing denuded exoplanets whilst the $\rm \sim 2.4R_{\oplus}$ radius peak mainly comprising exoplanets with large hydrogen envelops.
	\end{abstract}



\section{Introduction}

Ever since the discovery of the bimodal distribution in exoplanet sizes \citep{Fulton2017} with one peak at $\rm \sim 1.3 R_{\oplus}$, the other at $\rm \sim 2.4 R_{\oplus}$, and the minimum at $\rm \sim 1.75 R_{\oplus}$, multiple models have been put forward attempting to explain this shape. Although formational mechanisms and arguments have been proposed \citep[e.g.][]{Ginzburg2016,Gupta2018,Ginzburg2018,Zeng2018}, in this paper we will focus on how stellar XUV irradiation erodes away the primordial $\rm H_{2}-dominated$ atmospheres of exoplanets \citep[e.g.][]{Lecavelier2007, Ehrenreich2011, Lammer2013, Jin2014,Jin2018,Kubyshkina2018(2)} and its contribution in shaping the distribution. One of the most highly cited models is \citet{Owen2013, Owen2017} which assumes that Kepler planets orbiting Sun-like stars, have Earth-like interiors, and have masses that vary according to a Rayleigh distribution with a mode at $\rm \sim3M_{\oplus}$. According to their model, exoplanets with masses above $\rm \sim3M_{\oplus}$ had large $\rm H_{2}$ atmospheres whilst planets with masses beneath this threshold were born approximately bare. In this paper we take a different and more general approach to atmospheric mass-loss than \citet{Owen2013, Owen2017}. Specifically, we account for interiors that range from 100$\rm \: wt\%$ Fe - 100$\rm \: wt\%$ $\rm H_{2}O$ that host $\rm H_{2}$ envelops. These primordial envelops will scale from 0 $\rm \: wt\%$ (a negligible or non-existent atmosphere) to 100 $\rm \: wt\%$ (a negligible or non-existent core) of the total mass. We also do not focus on the process through which an exoplanet loses its atmosphere, just the beginning and the end result. By considering the multiplicity of compositions and envelops we evolve a synthetic population of $10^{6}$ exoplanets to test whether the bimodal distribution observed by \citet{Fulton2017} could be reproduced.

\section{Initial Conditions}

\subsection{Stellar Population}

As of September 2019 there are $\sim 4000$ known exoplanets in total, however for this study we are only interested in the Kepler candidates. To keep in line with \citet{Fulton2017} we only considered planets with orbital periods $\leq 100$ days. From these we could extract useful stellar information out of $2171$ via the \textit{NASA Exoplanet Archive}. Discarding O-, A-, B- and F0-F5 type stars (which were only 29 in number) left us with $2142$. The resultant distribution of stellar spectral types is shown in table~{\ref{tab:spectraltype}}. 

\begin{table}
	\centering
	\caption{Stellar spectral type frequency for known exoplanets with orbital periods $ \leq 100$ days. Data from the \textit{NASA Exoplanet Archive}.}
	\label{tab:spectraltype}
	\begin{tabular}{P{1.5cm}P{1.5cm}P{3.5cm}} 
		\hline
		\hline                     
		Type & Number & Norm. Fraction ($\%$) \\
		\hline
		M & 14 & 0.654 \\
		K & 376 & 17.554 \\
		G & 1311 & 61.204 \\
		F (F5-F9) & 441 & 20.588 \\
		\hline  
	\end{tabular}
\end{table}

Even if this data set is biased due to instrumental limitations, it was used to generate the bimodal distribution by \citet{Fulton2017}, so we may use it to generate a synthetic population of $10^{6}$ ZAMS stars with their spectral-type distribution corresponding to table~{\ref{tab:spectraltype}} and their masses, radii and temperatures matching those predicted by \citet{Siess2000}. We are aware that stars evolve with age but since most of the XUV-induced evaporation occurs within the first $\sim 1$ Gyr we can ignore this evolution and instead focus on the irradiation present during the star's youth. We also did not account for small variations within each spectral class of mass, radius, metallicity and temperature since the effects on our final results would be negligible.

\begin{table}
	\centering
	\caption{The properties of our synthetic population \citet{Siess2000}.}
	\label{tab:probability}
	\begin{tabular}{P{1.5cm}P{1.5cm}P{1.5cm}P{2.5cm}} 
		\hline
		\hline                     
		Type & Mass ($\rm M_{\ast}$) & Radius ($\rm R_{\ast}$) & Temperature (K) \\
		\hline
		M6 & 0.1 & 0.192 & 2973 \\
		M5 & 0.2 & 0.224 & 3244 \\
		M3 & 0.3 & 0.299 & 3474 \\
		M2 & 0.4 & 0.360 & 3654 \\
		M1 & 0.5 & 0.430 & 3811 \\
		M0 & 0.6 & 0.517 & 3982 \\
		K6 & 0.7 & 0.623 & 4261 \\
		K4 & 0.8 & 0.735 & 4617 \\
		K2 & 0.9 & 0.850 & 4925 \\
		K1 & 1.0 & 0.965 & 5239 \\
		G9 & 1.1 & 1.110 & 5475 \\
		G7 & 1.2 & 1.290 & 5685 \\
		G2 & 1.3 & 1.450 & 5917 \\
		F9 & 1.4 & 1.640 & 6116 \\
		F7 & 1.5 & 1.840 & 6296 \\
		F5 & 1.6 & 2.030 & 6489 \\
		\hline  
	\end{tabular}
\end{table}

In order to estimate the frequency of each sub-type we divided the frequency of each spectral class by the number of sub-types given by \citet{Siess2000}. Regarding the XUV stellar irradiation, we used the equations by \citet{Penz2008(1)} for M-type stars and \citet{Penz2008(2)} for G-type stars with the UV fluxes from \citet{Sanz-Forcada2011}. Since K-type stars are intermediate in X-ray luminosity levels we set their values as being in-between M- and G-types. For late F-type stars we set their XUV luminosities as equivalent to G0-type stars which is adequate for our statistical model.

\subsection{Orbital Parameters}

\begin{figure}[h]
	\centering
	\includegraphics[scale=0.80]{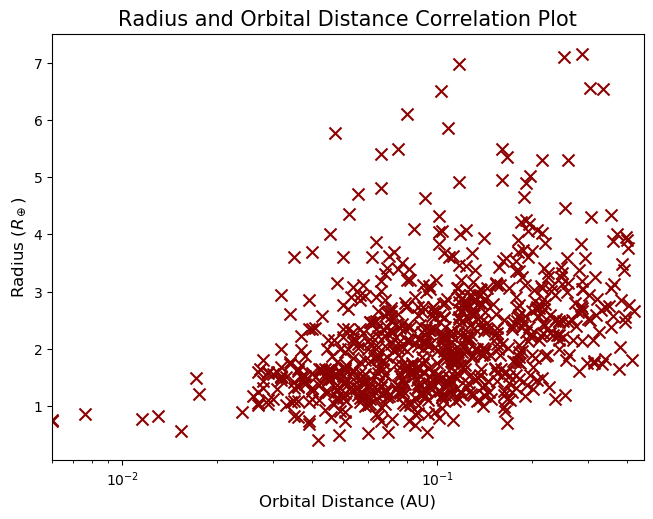}
	\caption{Semi-major axial distance (AU) plotted against planetary radius ($\rm R_{\oplus}$) for orbital periods $ \leq 100$ days. 787 data points all of which were collected on September 2019 from the \textit{NASA Exoplanet Archive}.}
	\label{fig:avsr}
	\centering
\end{figure}

\begin{figure}[h]
	\centering
	\includegraphics[scale=0.80]{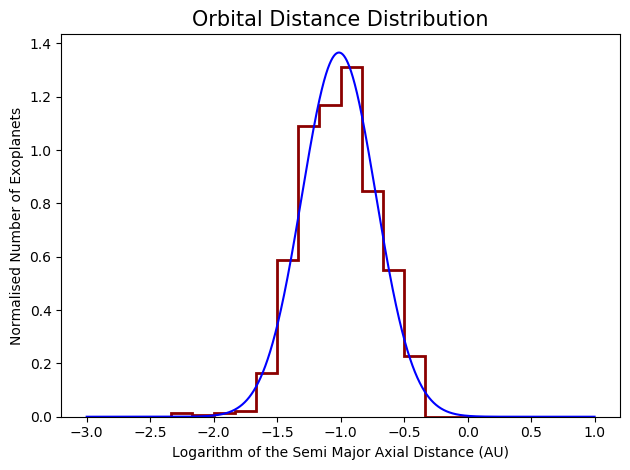}
	\caption{The normalised observed distribution of exoplanet orbital distances (red) and the lognormal approximation (blue) for exoplanets with orbital periods $ \leq 100$ days. 838 data points all of which were collected on September 2019 from the \textit{NASA Exoplanet Archive}.}
	\label{fig:adistribution}
	\centering
\end{figure}

When we plot the planetary radii versus the semi-major axis distances, a deficit of bodies intermediate in size between a super-Earth and a sub-Jupiter with short orbital periods (see figure~{\ref{fig:avsr}}) can be seen. This paucity is believed to be due to the photoevaporation of primordial atmospheres \citep[e.g.][]{Owen2018}. Since our code will evolve synthetic exoplanets from before the time that their atmospheres were eroded, we will assume that there was no gap straight after their formation so a planet with an arbitrary radius could have any arbitrary orbital distance. Notwithstanding, we set an upper limit for the radius of our synthetic exoplanets at $\rm R_{P} \approx 8.0R_{\oplus}$ (i.e. $\rm M \approx 25M_{\oplus}$ for a cold $\rm H_{2}$ body) in order to exclude Saturn- and Jupiter-mass planets which are more resilient to atmospheric erosion. With this assumption it became possible to begin sampling exoplanetary orbital distances. The histogram of the semi-major axial distances showed that the average orbit is at $\sim 0.1$ AU (see figure~{\ref{fig:adistribution}}). This distribution was best fit by a lognormal curve between a minimum distance set by the \textit{Roche limit} \citep{Aggarwal1976} of each individual star and a maximum distance corresponding to a period of 100 days. We approximated the \textit{Roche limits} of the stars as $\rm \sim 2R_{\ast}$, where $ R_{\ast}$ is the radius of the host star being considered at that moment, because planets with large $\rm H_{2}$ envelops such as the outer planets in our solar system have stellar-like densities. This approximation breaks down for exoplanets born with very small atmospheres but these should be scarce since most planets are thought to begin with large $\rm H_{2}$ repositories \citep[e.g.][]{Hayashi1985,Ikoma2012}.

\subsection{Planetary Parameters}
\label{sec:planetaryparameters}

\begin{figure}[h]
	\centering
	\includegraphics[scale=0.8]{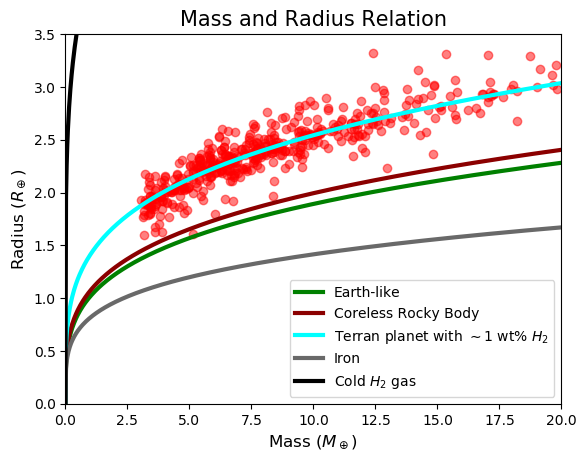}
	\caption{The mass and radius distribution of 500 of our synthetic exoplanets before undergoing photoevaporation. All mass and radius curves come from \citet{Zeng2013} and \citet{Zeng2016} except for the cold $\rm H_{2}$ curve which originated from the equations of state by \citet{Becker2014} that were then adapted by \citet{Zeng2018}. The radius distribution was given by a laplacian function with a mean corresponding to a terran planet with $\rm 1 \: wt\%$ $\rm H_{2}$ envelop and a scale parameter of $0.10$ (i.e. a standard deviation of $\sqrt{2}/10$). The reason for the lack of planets with $\rm M_{P} < 3M_{\oplus}$ is explained later in the manuscript.}
	\label{fig:massradius}
	\centering
\end{figure}

Our initial population of exoplanets was modelled as having $\rm H_{2}$ atmospheres so their average densities had to be lower than a purely silicate planet. However, it is possible that some exoplanets formed with very small or non-existent atmospheres whilst others formed almost entirely as hydrogen spheres with no core. Consequently we modelled the variation in exoplanet radii with a laplace distribution were for each planet the maximum allowed density was set as a 100$\rm \: wt\%$ Fe sphere, the minimum density was modelled as a 100$\rm \: wt\%$ $\rm H_{2}$ sphere, and the average density was given by a terran core with a $\rm H_{2}$-rich atmosphere. For the Fe and terran mass-radius models we approximated the numerical data in \citet{Zeng2013} and \citet{Zeng2016} with a best-fit curve whilst for the pure hydrogen planet we used \citet{Becker2014}. Figure~\ref{fig:massradius} shows the mass and radius relation of 500 synthetic exoplanets as given by our laplace distribution.
\\
\begin{subequations}
\\
100$\rm \: wt\%$ Fe planet radius \citep{Zeng2013,Zeng2016}:
	\begin{equation}
	\label{eq:Feplanet}
	\dfrac{R_{Fe}}{R_{\oplus}} = 0.815 \times \left(\dfrac{M_{P}}{M_{\oplus}} \right)^{1/4.176}
	\end{equation}    
Terran planet with $\rm \sim 1 \: wt\%$ $\rm H_{2}$-rich envelop radius \citep{Zeng2013,Zeng2016}:
	\begin{equation}
	\label{eq:rockplanet}
	\dfrac{R_{Rock+H_{2}}}{R_{\oplus}} = 1.410 \times \left(\dfrac{M_{P}}{M_{\oplus}} \right)^{1/3.905}
	\end{equation}
100$\rm \: wt\%$ $\rm H_{2}$ planet radius \citep{Becker2014,Zeng2018}:
	\begin{equation}
	\label{eq:gasplanet}
	\dfrac{R_{H_{2}}}{R_{\oplus}} = 4.106 \times \left(\dfrac{M_{P}}{M_{\oplus}} \right)^{1/5.010}
	\end{equation}
\end{subequations}

Where $M_{P}$ is the planetary mass in kg, $M_{\oplus}$ is Earth's mass in kg, and $R_{\oplus}$ is Earth's radius in m. Regarding the planetary mass, we have very few measurements for super-Earths and sub-Neptunes, and even when we do have values they sometimes suffer from large uncertainties. Because of this, we have very strong observational biases towards larger masses \citep[e.g.][]{Howard2010,Marcy2014,Malhotra2015} which means that as of now it is not possible to accurately and reliably know the mass function of exoplanets. Consequently, we will set the initial planetary mass distribution as a variable which we will manually adjust in order match the observed radius gap as given by \citet{Fulton2017}.

\begin{figure}[h]
	\centering
	\includegraphics[scale=0.80]{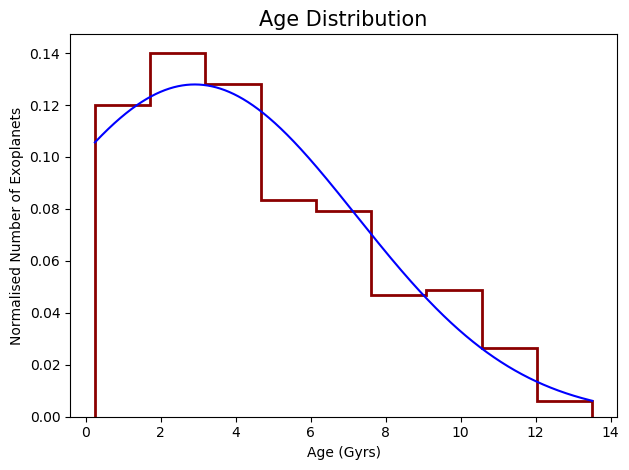}
	\caption{The observed normalised distribution of exoplanet ages (red) and the truncated gaussian approximation (blue) for orbital periods $ \leq 100$ days. All 341 data points were collected on September 2019 from the \textit{NASA Exoplanet Archive}. We only used Kepler candidates.}
	\label{fig:tdistribution}
	\centering
\end{figure}

Some planets are young enough that their atmospheres are still extant despite strong irradiation that would, in the future, denude the planet. These bodies are younger than the amount of time required for their complete atmospheric loss. Using the age of their host stars which is given by the \textit{NASA Exoplanet Archive} as proxies, the age distribution of exoplanets is best described by a truncated gaussian with a mean of $\sim 2.9 $ Gyr, standard deviation of $\sim 4.3 $ Gyr and a minimum and maximum age of $\sim 0.25 $ Gyr and $\sim 13.5 $ Gyr respectively (shown in Figure~\ref{fig:tdistribution}). It is important to note that the age of the stars cannot be directly measured and it is calculated using theoretical modelling which is known to be uncertain. However, it is beyond the scope of this study to do an in-depth analysis of these models.
With regards to the mass-loss timescale, for a super-Earth or sub-Neptune this is approximated by:
\begin{subequations}
\begin{equation}
\label{eq:masslosstimescale}
t_{mass\: loss} \sim \dfrac{M_{env} }{\dot{M}_{env}}
\end{equation}
Where
\begin{equation}
\label{eq:envelopmass}
M_{env} = \begin{cases} 0.01M_{P} &\text{if $M_{P} < \Upsilon_{\alpha}$} \\ M_{P} - \Gamma_{\beta} &\text{if $\Upsilon_{\alpha} \leq M_{P} < \Upsilon_{\beta}$} \\ M_{P} - \Gamma_{\gamma} &\text{if $M_{P} \geq \Upsilon_{\beta}$} \end{cases}
\end{equation}
\end{subequations}

Where $M_{env}$ is the mass of the primordial envelop, $\dot{M}_{env}$ is the hydro-based mass-loss rate which is described in \citet{Kubyshkina2018(1)}. Since in our study we only focused on the beginning and end result of the mass-loss evolution, we could not include the time dependence on the XUV flux rate. Therefore, in order not to make an overestimation (e.g. by assuming a constant initial flux throughout the lifetime of the star), we modelled this parameter as a step function where $\rm F_{XUV} = F_{0}$ from $0-1$ Gyrs and $\rm F_{XUV} = 0$ for $>1$ Gyrs. The initial flux $F_{0}$ was adopted from \citep{Penz2008(1),Penz2008(2),Sanz-Forcada2011}. In addition, the treatment of the flux as a step function is consistent with these studies as during the first billion years of a stars' life the XUV flux is approximately constant, with it then quickly decaying by several orders of magnitude. In the appendix we present Figures~(\ref{fig:gtypexuv},\ref{fig:mtypexuv}) which show that this approximation is adequate. Regarding $\Upsilon_{\alpha}$ and $\Upsilon_{\beta}$, these are the critical masses that separate the different regimes $\alpha$, $\beta$ and $\gamma$. $\Gamma_{\beta}$ and $\Gamma_{\gamma}$ are probabilistic functions with ranges between $\rm \Upsilon_{\alpha}-M_{P}$ and $\rm \Upsilon_{\beta}-M_{P}$ respectively. When $\rm \Gamma_{\beta \: \text{or} \: \gamma} = M_{P}$ this implies that the planetary envelop is nonexistent, whilst if $\rm \Gamma_{\beta} = \Upsilon_{\alpha}$ or $\Gamma_{\gamma} = \Upsilon_{\beta}$ are for the cases when the planet has the minimum core mass required to be in that specific region:
\begin{enumerate}
	\item $\alpha$ marks the region where smaller planets typically accrete hydrogen envelops that are $\sim 1\%$ of the total planetary mass \citep[e.g.][]{Stevenson1999,Ikoma2012,Chachan2018}. This region will be within the mass limits of $0 - \Upsilon_{\alpha}$.
	\item $\beta$ is where planets have enough mass to accrete larger envelops but the cores are still not massive enough to form Neptunian- or Jupiter-mass bodies \citep[e.g.][]{Ikoma2012,Chachan2018}. Region $\beta$ will be within the mass limits of $\Upsilon_{\alpha}$ and $\Upsilon_{\beta}$.
	\item In the $\gamma$ region planets undergo runaway gas accretion. This implies that generally any extra mass above the critical mass $\Upsilon_{\beta}$ will mostly be due to hydrogen gas. Generally theoretical models predict that the critical mass $\Upsilon_{\beta} \rm \sim 10M_{\oplus}$ \citep[e.g.][]{Ida2004,Ida2005,Ida2008,Mordasini2009}. However, as systems undergo runaway growth planetesimals are also accreted which could increase the mass of the core \citep{Shiraishi2008,Shibata2019}. This extra growth of the core needs to be balanced with the core erosion induced from deposited energy but the latter effect is believed to be inefficient \citep{Moll2017}. In addition, internal models of the gaseous planets in our solar system predict core masses in the range of $\rm 7-25M_{\oplus} $ \citep[e.g.][]{Mizuno1980,Stevenson1982,Hubbard1989,Chabrier1992,Guillot1997,Gudkova1999,Guillot1999,Wahl2017}. Therefore the remnant cores of Neptunian or Jovian planets could be $\rm >10 M_{\oplus}$. The possibility of very heavy cores is also supported by the discovery of rocky mega-Earths such as BD+20 594 b which has a mass and radius of $\rm 16.3 \pm 6.0 M_{\oplus} $ and $\rm 2.2 \pm 0.1R_{\oplus} $ \citep{Espinoza2016} or K2-66 b with $\rm 21.3 \pm 3.6M_{\oplus} $ and $\rm 2.50 \pm 0.3R_{\oplus} $ \citep{Sinukoff2017} respectively. Therefore whilst the critical mass $\rm \Upsilon_{\beta}\sim 10M_{\oplus}$, the probabilistic function $\rm \Gamma_{\gamma}$ could range from $\rm 10-25M_{\oplus}$ according to observations and theoretical predictions.
\end{enumerate}
Having considered all of the above, we will adjust our input distributions for $ \Gamma_{\beta} $ and $ \Gamma_{\gamma} $ and our critical masses $\Upsilon_{\alpha}$ and $\Upsilon_{\beta}$ in order to find a best fit radius distribution.

\section{Evolution Model}

The simplest summary of our model is that our synthetic exoplanets travel one of three evolutionary paths:

\begin{enumerate}
	\item The exoplanet is younger than its mass-loss timescale so it currently maintains its envelop.
	\item The exoplanet has the right orbital and physical properties required for its primordial envelop to outlive its host star's lifetime \citep{Locci2019}.
	\item The exoplanet loses its atmosphere before undergoing any counter-mechanism.
\end{enumerate}

\begin{figure}[h]
	\centering
	\includegraphics[scale=0.75]{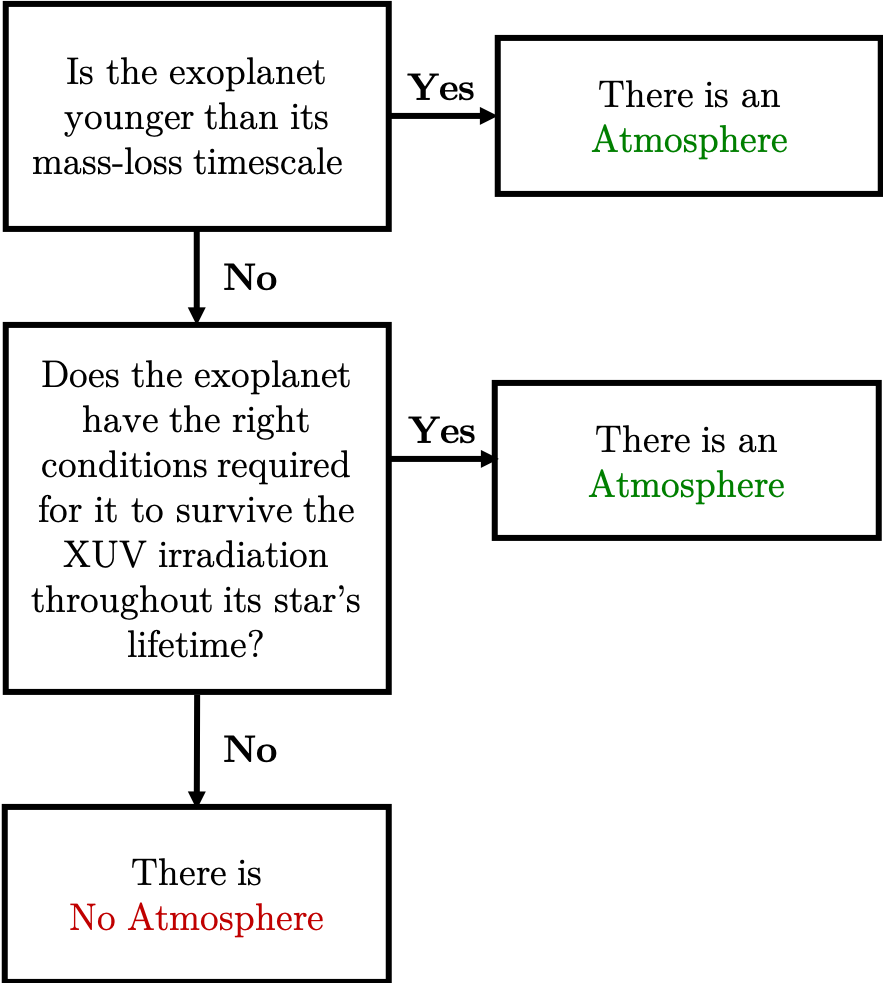}
	\caption{Flowchart showing the path our code took when evolving the exoplanets.}
	\label{fig:flowchart}
	\centering
\end{figure}

In our code we integrated an algorithm (see figure~{\ref{fig:flowchart}}) which implemented the above points. Now that we knew each evolutionary path, we needed to model each step accordingly.

\subsection{First Path}

If the age of the planet is less than the mass loss timescale we did not evolve it (see equation 3). However, if the contrary is true we moved it to the next stage.

\begin{equation}
\label{eq:inequality}
t_{age} < t_{mass \medspace loss}
\end{equation}

\subsection{Second Path}

The second path is concerned with whether the initial mass of an exoplanet's envelop is large enough to survive the incident XUV irradiation for the remainder of the host star’s life. In order to estimate this minimum survivable mass, we first calculated the mass loss rate for different synthetic giant exoplanets. To do this  we adapted the numerical model for the mass loss rate described in \citet{Locci2019}. This model calculates the mass loss rate using the “energy-limited approach” which was first proposed by \citet{Watson1981} and was later revisited by \citet{Erkaev2007}. We then performed a temporal evolution of the synthetic gaseous  planet  population, where at each arbitrary unit of time we:
\begin{enumerate}
	\item Calculated the  mass loss rate.
	\item Updated the  total planetary mass.
	\item Updated  the planetary radius accounting for both: mass lost and gravitational shrinking \citep[see ][for details]{Locci2019}.
\end{enumerate}
For the duration of the simulation we adopted:
\begin{itemize}
	\item An average G-type lifetime of 10 Gyrs.
	\item Different possible exoplanet orbital distances.
	\item Varying initial X-ray luminosities of the parent star.
\end{itemize}
After the evolution we took note of the planets that ended up as super-Earths from which we then retrieved their initial masses for their given distances and stellar X-ray luminosities. When a planet ended its evolution with a mass less than $\rm 2.6 M_{\oplus}$, we defined it as having lost all of its gaseous envelope. We chose $\rm 2.6 M_{\oplus}$ as according to the mass-radius relations of \citet{Zeng2013} and \citet{Zeng2016} this would correspond to a rocky planet with a radius at $\rm \sim 1.3R_{\oplus}$ which according to \citet{Fulton2017} is located at the first peak of the radius distribution. If the planet instead has a hydrogen envelop with a total combined mass of $\rm \rm 2.6 M_{\oplus}$, its radius would be $\rm \sim 1.8R_{\oplus}$ which is located within the Fulton gap. Consequently, this mass threshold implies that planets originally located within the region of the radius gap, lost their envelops to `fill-up' the first peak at $\rm \sim 1.3 R_{\oplus}$.

\begin{figure}[h]
	\centering
	\includegraphics[scale=0.7]{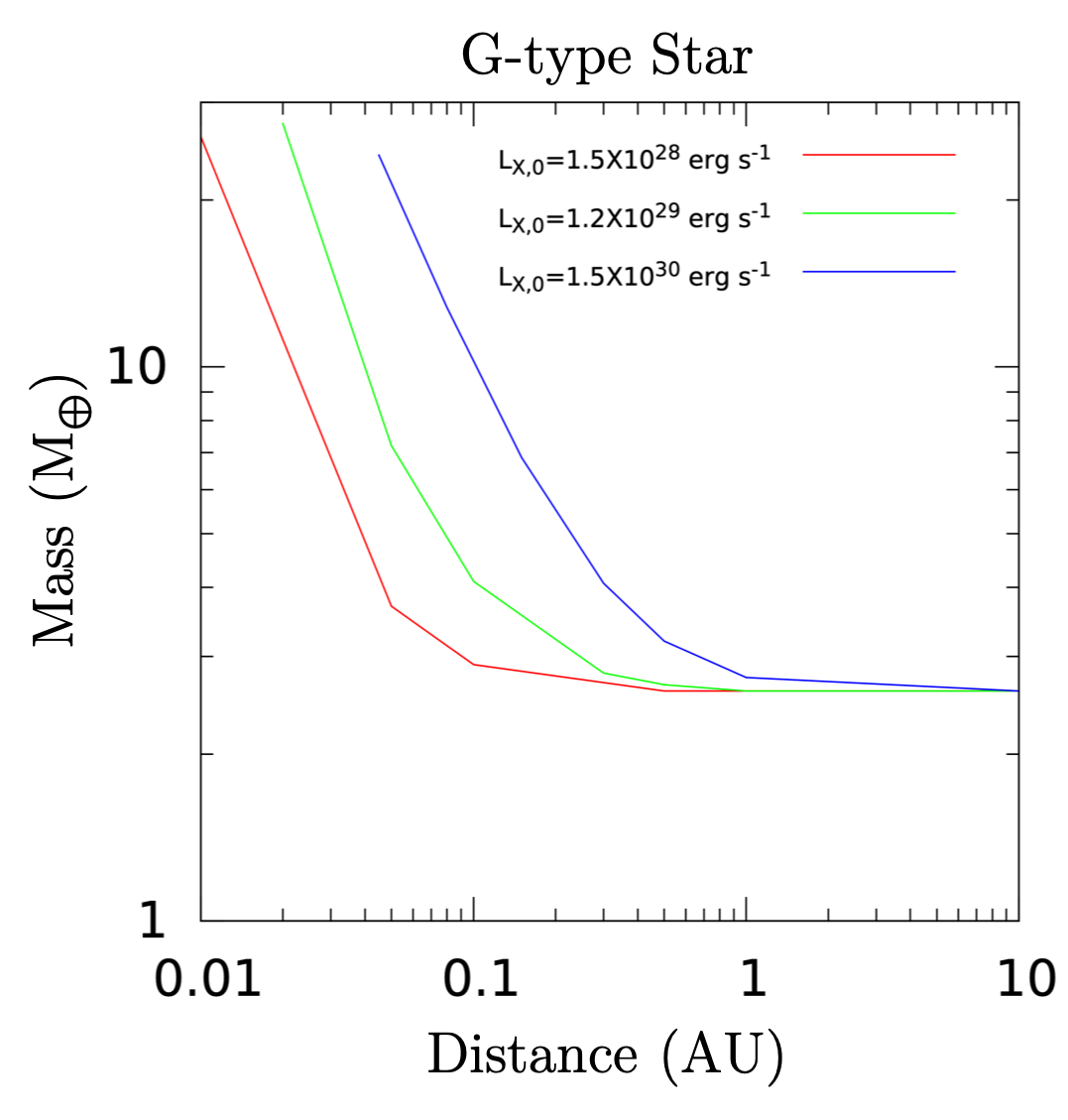}
	\caption{The mass threshold functions calculated using the model from \citet{Locci2019} for G-type stars for different initial X-ray luminosities \citep{Penz2008(2)}. The UV luminosities scale with the X-ray ones according to the models by \citet{Sanz-Forcada2011}.}
	\label{fig:gtype}
	\centering
\end{figure}

\begin{figure}[h]
	\centering
	\includegraphics[scale=0.7]{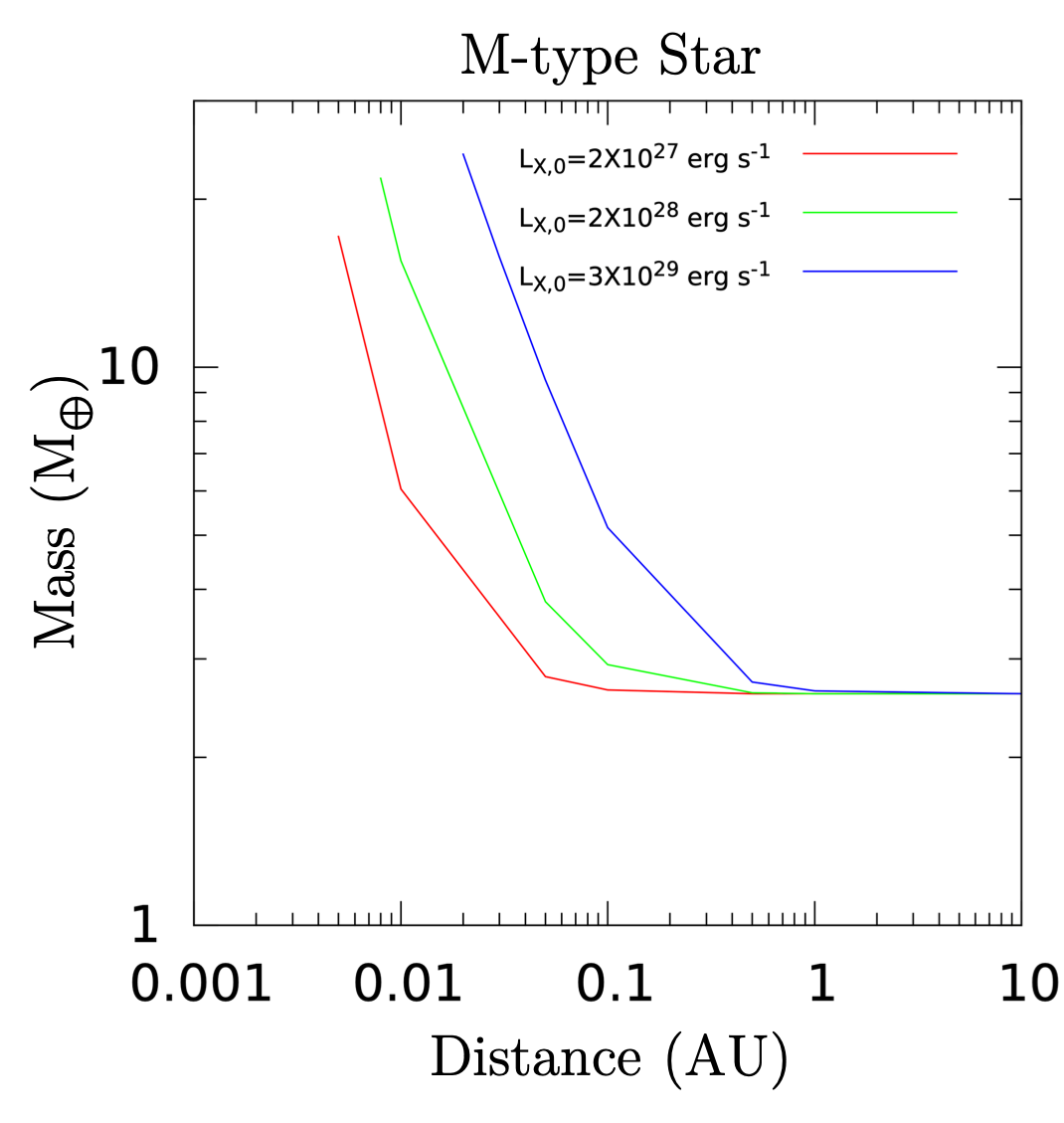}
	\caption{The mass threshold functions calculated using the model from \citet{Locci2019} for M-type stars for different initial X-ray luminosities \citep{Penz2008(1)}. The UV luminosities scale with the X-ray ones according to the models by \citet{Sanz-Forcada2011}.}
	\label{fig:mtype}
	\centering
\end{figure}

We also accounted for the fact that luminosities evolve with time as described in \citet{Penz2008(2)} and \citet{Penz2008(1)}. Finally, using this procedure, we retrieved the initial mass thresholds as functions of the orbital distance, for different stellar luminosities, and for M- and G-type stars (in figure \ref{fig:gtype} and \ref{fig:mtype} we show the mass threshold functions for a G-type and M-type star respectively for given initial X-ray luminosities). These are the mass limits required for an exoplanet with a certain mass and radius to survive its host star’s XUV irradiation throughout its lifetime. During our exoplanet simulations around their host stars we occasionally interpolated between the calculated values of the minimum mass in order to obtain a best-fit curve for each given luminosity.

\subsection{Third Path}

\begin{figure}[h]
	\centering
	\includegraphics[scale=0.8]{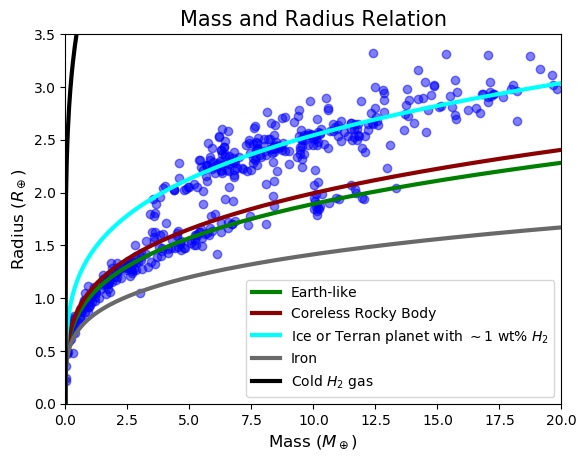}
	\caption{The mass and radius distribution of 500 of our synthetic exoplanets after undergoing photoevaporation. All mass and radius curves come from \citet{Zeng2013} and \citet{Zeng2016} except for the cold $\rm H_{2}$ curve which originated from the equations of state by \citet{Becker2014} that were then adapted by \citet{Zeng2018}. The scale parameter of the laplacian function of the radii was set to $0.05$ (i.e. a standard deviation of $\sqrt{2}/20$.)}
	\label{fig:massradius2}
	\centering
\end{figure}

If an exoplanet makes it to the third stage, it will lose its hydrogen atmosphere. In order to model this, we subtracted the atmospheric mass (given in Equation~\ref{eq:envelopmass}) from the total planetary mass ($M_{P}$). Once the masses had been reduced we then calculated their corresponding radii. Since these planets would lack hydrogen atmospheres, their minimum possible densities are consistent with $100\%$ $\rm H_{2}O$ ice planets \citep{Zeng2013,Zeng2016}, whilst their maximum possible densities are given by the remnant cores of evaporated Neptunian or Jovian planets \citep{Mocquet2014}. The mean density of our evolved planets was set to an Earth-like silicate planet with a core comprising $30\%$ of the total mass \citep{Zeng2016}. We distributed the occurrence of each of these compositions according to a laplacian distribution which is shown in Figure~\ref{fig:massradius2}.\\

\begin{subequations}
Gas giant or Neptunian remnant core \citep{Mocquet2014}:
	\begin{equation}
	\label{eq:remnantcore}
	\dfrac{R_{Remnant \: Core}}{R_{\oplus}} = 0.469 \times \left(\dfrac{M_{P}}{M_{\oplus}} \right)^{1/3}
	\end{equation}    
Terran planet with no atmosphere \citep{Zeng2013,Zeng2016}:
	\begin{equation}
	\label{eq:terranplanet}
	\dfrac{R_{Rock}}{R_{\oplus}} = 1.007 \times \left(\dfrac{M_{P}}{M_{\oplus}} \right)^{1/3.7}
	\end{equation}
100$\%$ $\rm H_{2}O$ planet radius \citep{Zeng2013,Zeng2016}:
	\begin{equation}
	\label{eq:iceplanet}
	\dfrac{R_{H_{2}O}}{R_{\oplus}} = 1.410 \times \left(\dfrac{M_{P}}{M_{\oplus}} \right)^{1/3.905}
	\end{equation}
\end{subequations}
However, if the original (pre-evolution) planetary density was already higher than an Earth-like silicate planet, we also sampled its new radius with a laplace distribution but between the limits set by its pre-evolution radius and the radius of a remnant gas giant/Neptunian core. The mean of the distribution was also set at its pre-evolution radius (see Table~\ref{tab:inputdistribution} for more details). Nevertheless, the majority of newly-formed planets are expected to have densities lower than purely silicate bodies due their primordial envelops. Consequently, the contribution from these rare exoplanets is expected to be negligible.

Comparing Figure~\ref{fig:massradius2} with \ref{fig:massradius} shows the expected XUV-irradiation-induced evolution in the mass-radius curves of exoplanets. Whilst the original mass and radius plot is centred around the ice or rocky planet with $\rm H_{2}$ envelop line, the second plot shows signs of two separate compositions. These being the original low-mass low-density planets, and rocky worlds. A small gap can also be distinguished between the ice/terran planet with a $\rm H_{2}$ envelop line and the Earth-like composition line. Figure~\ref{fig:massradius2} follows a similar trend to the mass and radius plot of real exoplanets as shown in Figure 2 of \citet{Zeng2018}.

\section{Results}

\begin{figure}[h]
	\centering
	\includegraphics[scale=0.85]{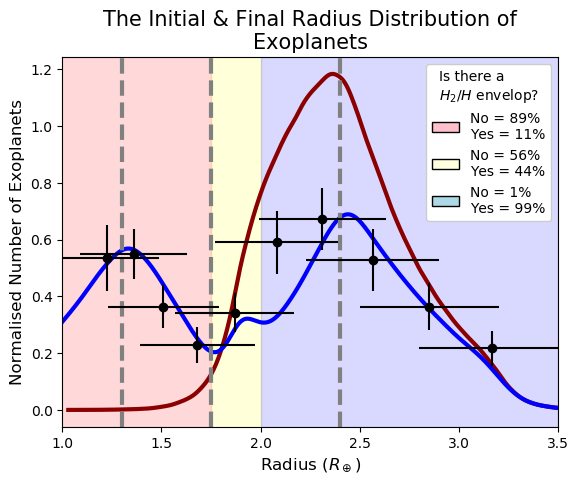}
	\caption{The final (blue line) and initial (red line) distributions of exoplanet radii for periods $ \leq 100$ days obtained by our model. The grey lines show the two peaks and the minimum discovered by \citet{Fulton2017}. The data points with the uncertainties correspond to the values given in Table 3 and Figure 7 of \citet{Fulton2017}. The light-red, light-yellow and light-blue regions correspond only to the final radius distribution's rocky first peak, the remnant core (mega-Earth)-rich region, and the hydrogen-rich worlds of the second peak respectively. For each region we report the planet population composition.}
	\label{fig:bimodaldistribution}
	\centering
\end{figure}

\begin{figure}[h]
	\centering
	\includegraphics[scale=0.8]{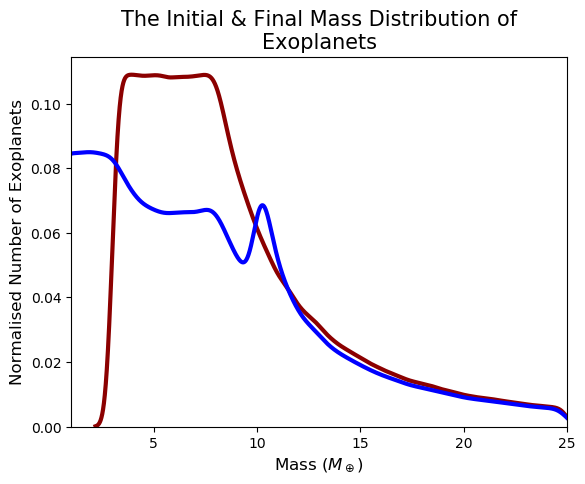}
	\caption{The predicted final (blue) and initial (red) mass distributions of exoplanets whose current radii are distributed according to Fig~{\ref{fig:bimodaldistribution}}. All planets have periods $ \leq 100$ days.}
	\label{fig:finalmassdistribution}
	\centering
\end{figure}

After evolving $10^{6}$ exoplanets we are able to replicate the observed bimodal distribution in exoplanet radii with one peak at $\rm \sim 1.3 R_{\oplus}$, the other at $\rm \sim 2.4 R_{\oplus}$, and the minimum at $\rm \sim 1.75 R_{\oplus}$ (see Figure~\ref{fig:bimodaldistribution}). Our results matched well with the observations by \citep{Fulton2017} whilst also being consistent with the presence of the sub-Jovian desert \citep[e.g.][]{Owen2018} (see Figures~\ref{fig:avsr} and \ref{fig:subjoviandesert}). Furthermore, our final radius distribution is within the error bars given by the data points in Table 3 and Figure 7 of \citet{Fulton2017}. Our model predicts that initially most exoplanets were centred around a peak of $\rm \sim 2.4 R_{\oplus}$ that decayed in a laplacian manner as shown in Figure~\ref{fig:bimodaldistribution}. In order to achieve the bimodal behaviour we found that it was mandatory for there to be a paucity of exoplanets $\rm \gtrsim 8 M_{\oplus}$ (see figure~\ref{fig:finalmassdistribution}). For best results we found that this region should be preceeded by a flat distribution between $\rm 3 - 8 M_{\oplus}$ and another paucity for planets $\rm \lesssim 3M_{\oplus}$. Because our best-fit initial mass distribution has a lack of planets with $\rm \lesssim 3M_{\oplus}$, we predict that within the confines of our observations, bodies situated in section $\alpha$ of Equation~\ref{eq:envelopmass} should be absent or in the very least relatively scarce. With respect to the initial hydrogen envelop masses in Equation~\ref{eq:envelopmass}, we find best-fit critical masses of $\rm \Upsilon_{\alpha}=3M_{\oplus}$, $\rm \Upsilon_{\beta}= 10M_{\oplus}$ with the probabilistic functions $\Gamma_{\beta}$ and $\Gamma_{\gamma}$ detailed in Table~\ref{tab:inputdistribution}.

\begin{figure}[h]
	\centering
	\includegraphics[scale=0.8]{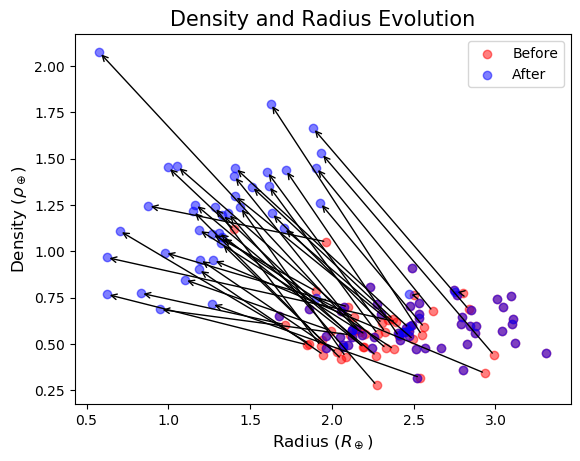}
	\caption{A plot showing the density and radius evolution of 100 randomly chosen exoplanets.}
	\label{fig:densityevolution}
	\centering
\end{figure}

\begin{figure}[h]
	\centering
	\includegraphics[scale=0.8]{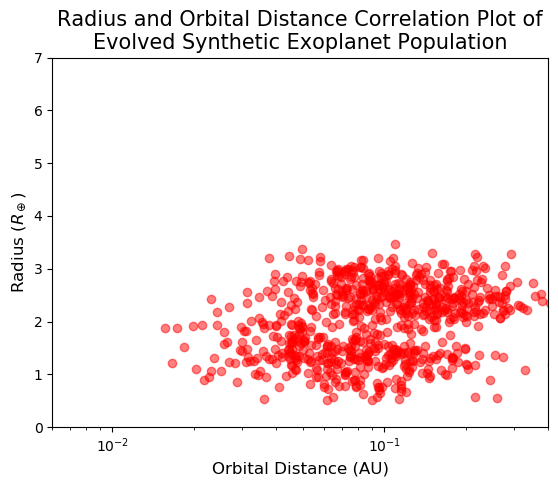}
	\caption{A plot of the radius and orbital distance of 787 (for comparison with Figure~\ref{fig:avsr}) randomly chosen exoplanets.}
	\label{fig:subjoviandesert}
	\centering
\end{figure}

Regarding the final mass distribution (see Figure~\ref{fig:finalmassdistribution}), for masses greater than $\rm \sim 11.5M_{\oplus}$ the difference between the initial and final distribution is very small. This is because with bigger masses it becomes exceedingly harder to remove the primordial hydrogen envelops of exoplanets due to their stronger gravitational strengths. From $\rm 10-11.5M_{\oplus}$ we predict a small increase in the abundance which is caused by the remnant cores of Neptunian or Jovian planets (more details on this can be found in section~\ref{sec:discussions}). From $\rm 3 - 10 M_{\oplus}$ there is a strong decrease in the mass abundance which is caused by planets with cores $\rm \lesssim 3M_{\oplus}$ losing the entirety of their envelops. After the XUV-induced evolution of these planets, the increase in the number of bodies with $\rm \lesssim 3M_{\oplus}$ corresponds with the radius peak at $\rm \sim 1.3R_{\oplus}$ \citep{Zeng2013,Zeng2016}. In spite of the increase in rocky worlds with masses at $\rm \sim 10M_{\oplus}$ matching the small peak at $\rm 1.95R_{\oplus}$ (despite the uncertainties being large, this is consistent with the $\rm 1.77-1.97R_{\oplus}$ abundance measured by \citet{Fulton2017}), the overall number of planets with radii of $\rm \sim 1.95R_{\oplus}$ has decreased. This is because despite the increase in rocky cores at $\rm \sim 1.95R_{\oplus}$, planets with hydrogen envelops previously located in this region decreased more numerously. Concerning with the radius peak at $\rm \sim 2.4R_{\oplus}$, this is compatible with rocky planets sustaining hydrogen envelops where the combined total mass is $\rm \sim 8M_{\oplus}$. According to \citet{Zeng2013,Zeng2016} An airless rocky planet with the same mass would instead be located at $\rm \sim 1.9R_{\oplus}$ which lies in the region of the remnant Neptunian and Jovian cores explained above. With respect to the density evolution our simulation shows that on average the planet densities increased sharply which can be seen in Figure~\ref{fig:densityevolution}. Numerically, our results can be summarised as follows with the best-fit input distributions shown in Table~\ref{tab:inputdistribution}:

\begin{enumerate}
	\item $\sim 1 \%$ of known exoplanets are currently in the processes of losing their atmospheres (first path).
	\item $\sim 51\%$ of exoplanets have the right orbital and physical properties required in order for their primordial envelops to outlive their host star's lifetime (second path).
	\item $\sim 48 \%$ of exoplanets completely lost their primordial hydrogen envelops (third path).
\end{enumerate}

\begin{table}
	\centering
	\caption{Adopted Input Distributions.}
	\label{tab:inputdistribution}
	\begin{tabular}{P{2.5cm}P{6cm}P{4cm}} 
		\hline
		\hline                     
		Exoplanet Property & Functional Form & Parameters \\
		\hline
		Orbital Distance & Lognormal & \begin{itemize} \item Max = 100 days \item Min = $\rm R_{Roche} \sim 2R_{\ast} $ \item  $\mu \approx$ -1 \item $\sigma \approx$ 0.3 \end{itemize} \\
		Initial Mass & $\rm = \begin{cases} \text{Uniform} & 3M_{\oplus} \leq M_{P} < 8M_{\oplus} \\ \text{Pareto} & 8M_{\oplus} \leq M_{P} \leq 25M_{\oplus} \end{cases}$ & \underline{Pareto} \begin{itemize} \item  $\rm Mo = 8.0 M_{\oplus}$ \item $\rm a = 1.6$ \end{itemize} \\
		Initial Compositions & Laplace & \begin{itemize} \item Max = Eq.~\ref{eq:gasplanet} \item Min = Eq.~\ref{eq:Feplanet} \item  $\mu = $ Eq.~\ref{eq:rockplanet} \item $\rm b = \sqrt{2}\sigma = 0.1 R_{\oplus}$ \end{itemize} \\
		Ages & Trunc. Normal & \begin{itemize} \item Max = $13.51$ Gyr \item Min = $0.25$ Gyr \item  $\mu \approx 2.9 $ Gyr \item $\rm \sigma = 4.3$ Gyr \end{itemize} \\
		$\Gamma_{\beta}$ (Eq.~\ref{eq:envelopmass}) & Uniform & \begin{itemize} \item Max = $\rm M_{P}$ \item Min = $0$ \end{itemize} \\
		$\Gamma_{\gamma}$ (Eq.~\ref{eq:envelopmass}) & Pareto & \begin{itemize} \item Max = $\rm M_{P}$ \item Min = $0$ \item Mo = $\rm 10M_{\oplus}$ \item a = $1$ \end{itemize} \\
		Final Compositions & Laplace & \underline{$\rm R_{Initial} > $ Eq.~\ref{eq:terranplanet}} \begin{itemize} \item Max = Eq.~\ref{eq:iceplanet} \item Min = Eq.~\ref{eq:remnantcore} \item  $\rm \mu =$ Eq.~\ref{eq:terranplanet} \item $\rm b = \sqrt{2}\sigma = 0.05 R_{\oplus}$ \end{itemize} \underline{$\rm R_{Initial} < $ Eq.~\ref{eq:terranplanet}} \begin{itemize} \item Max = $\rm R_{Initial}$ \item Min = Eq.~\ref{eq:remnantcore} \item  $\rm \mu = R_{Initial}$ \item $\rm b = \sqrt{2}\sigma = 0.05 R_{\oplus}$ \end{itemize} \\
		\hline  
	\end{tabular}
\tablecomments{Mo = Mode, $\mu$ = arithmetic mean, $\sigma$ = standard deviation, $a$ = shape, $b$ = scale, Max and Min are the maximum and minimum limits respectively.}
\end{table}

\section{Discussions}
\label{sec:discussions}

Overall our results are similar to those of \citet{Owen2013,Owen2017} where the observed radial distribution was caused due to stellar XUV irradiation triggering atmospheric erosion. However, the differences are due to us adopting the core compositional models of \citet{Zeng2013,Zeng2016} and \citet{Mocquet2014} which use considerably different equations of state. Our code also predicts a small but non-negligible contribution from large remnant cores (i.e. mega-Earths) formed from the removal of large hydrogen atmospheres from Neptunian-planets. We argue that most of these remnant cores are of mass $\rm \sim 10M_{\oplus}$ which for decompressed relaxed cores (which occurs after $\rm \sim few \; Gyrs$) corresponds to radii within the range $\rm 1.77-1.97R_{\oplus}$. This result is compatible with \citet{Swain2019} that indentified a ``transition region'' rich in terrestrial and hydrogen-bearing planets between $\rm 1.5R_{\oplus}-2.0R_{\oplus}$. As explained in section \ref{sec:planetaryparameters} some remnant cores may be much larger which is corroborated by observational evidence and a strong theoretical foundation. These larger mega-Earths have radii that are located within the $\rm \sim 2.4R_{\oplus}$ peak implying that this larger peak is dominated by planets with large hydrogen envelops ($\sim 99\%$) with a few massive mega-Earths ($\sim 1\%$).

Regarding the composition of each peak, to a very tight standard deviation (see Figures~\ref{fig:massradius} and \ref{fig:massradius2} for a visual representation, and Table~\ref{tab:inputdistribution} for the numerical values) we find that most planets in the first peak ($\rm \sim 1.3 R_{\oplus}$) are rocky and without a primordial atmosphere which agrees with the results from \citet{Owen2013,Owen2017}, \citet{Jin2018} and \citet{Swain2019}. Conversely, we find that most planets in the second peak ($\rm \sim 2.4 R_{\oplus}$) have large primordial envelops. Notwithstanding, there are exceptions such as Fe planets with small hydrogen atmospheres that could lay in the first peak or denuded $\rm H_{2}O$ ice planets that have very low densities and therefore may lay in the second peak \citep{Zeng2018} but these are very scarce. We believe that our predicted initial distribution of exoplanet radii (figure~{\ref{fig:bimodaldistribution}}) makes conceptual sense as most planets are born with large hydrogen depositories which results in their large puffy radii. Of these planets, very few have unusually small radii (i.e. small atmospheres) or unusually large radii (i.e. large atmospheres) which results in the initial distribution looking laplacian in nature. 

\begin{figure}[h]
	\centering
	\includegraphics[scale=0.75]{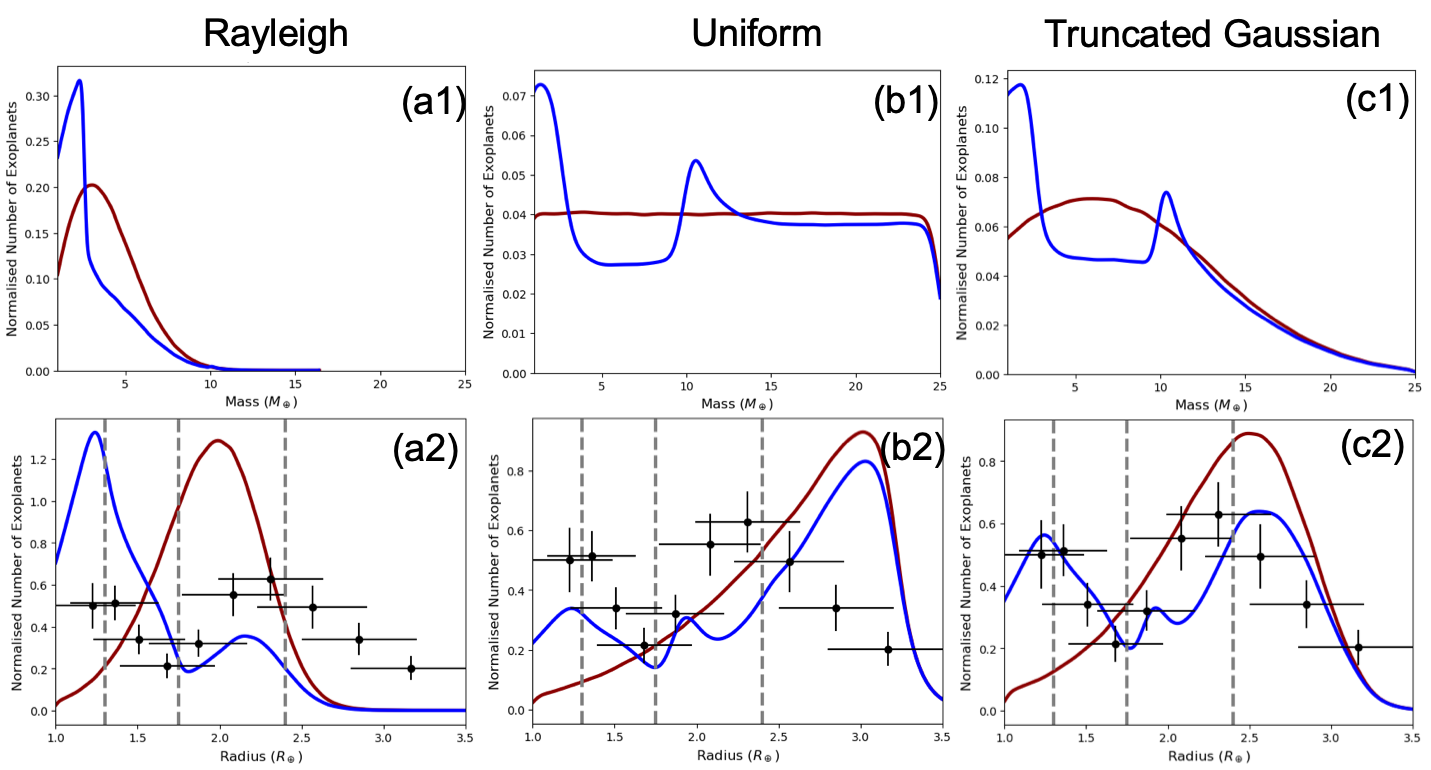}
	\caption{A plot showing the initial (red) and final (blue) mass (a1, b1, c1) and radius (a2, b2, c2) functions. The truncated gaussian function for the mass distribution (c1) has a mean of $\rm 6M_{\oplus}$, a standard deviation of $\rm 7M_{\oplus}$, and a minimum and maximum limit of $\rm 0.1$ and $\rm 25M_{\oplus}$ respectively.}
	\label{fig:collage}
	\centering
\end{figure}

One major difference between our paper and \citet{Owen2017} is our predicted initial mass distributions of exoplanets. In their paper, their mass function follows a Rayleigh distribution with a mode at $\rm 3 M_{\oplus}$. When we adopted the same distribution we were unable to generate the bimodal behaviour with the right amplitudes at each peak (see Figure~\ref{fig:collage}). This disparity is explained by the intrinsic nature of Rayleigh distributions which generally produce a large number of values relatively close to zero. According to our simulations in order to achieve a distribution emulating the one by \citet{Fulton2017} it is necessary for there to be a deficiency in planets with small masses so a Rayleigh function is not appropriate. In addition, our best-fit initial mass distribution requires a lack of exoplanets with masses $\rm \gtrsim 8 M_{\oplus}$. According to our simulations, this lack of Neptunian-mass bodies is independent of stellar XUV irradiation. This drop in the occurrence rate of larger mass exoplanets is consistent with observations \citep[e.g.][]{Howard2010,Marcy2014,Malhotra2015}. In addition, a paucity in exoplanet masses beyond $\rm \sim 10M_{\oplus}$ has been predicted by several planetary formation models \citep[e.g.][]{Ida2004,Ida2005,Ida2008,Mordasini2009} as at this critical size efficient accretion of materials results in few bodies having intermediate masses between $\rm 10-100M_{\oplus}$. Concerning with the flat distribution between $\rm 3-8 M_{\oplus}$, this was not mandatory for a bimodal distribution in the radii, but it gave us a better fit than our best-fit truncated gaussian function (see Figure~\ref{fig:collage}). There are several processes that could give rise to this result. One potential explanation is that even though small cores are more common than larger ones \citep[e.g.][]{Schlichting2013,Simon2016}, they are also harder to detect. Planets with smaller masses not only accrete less primordial gas \citep[e.g.][]{Stevenson1999,Ikoma2012,Chachan2018}, but they are also more susceptible to atmospheric destruction. In addition, they are also more likely to be ejected from their planetary systems because of gravitational pertubations. This is particularly true if giant planets are present in the system, with simulations predicting that on average each star ejects $\sim 7.9$ planets of which $\sim 2.5$ are terrestrial in size \citep{Barclay2017}. Although the exact demographics of rogue planets are a subject of dispute, it is generally accepted that there are at least billions in the Milky Way galaxy \citep[e.g.][]{Levison1998,Chambers1998,Debes2007}. Furthermore, at very close distances to the host star (i.e. $\lesssim 1$ AU) there is a scarcity of solids which limits the growth of planetesimals \citep[e.g.][]{Lodders2003,Mordasini2009}. Consequently, most seed embryos can only form beyond the ice line \citep[e.g.][]{Podolak2004,Martin2012,Dangelo2015}. However, fast type 1 migration means that the depletion in planets with $\rm 3M_{\oplus} \lesssim M_{P} \lesssim 10M_{\oplus}$ disappears because bodies initially orbiting farther out move inwards \citep{Mordasini2009}. This implies that super-Earths and sub-Neptunes with masses $\rm \lesssim 3M_{\oplus}$ would be preferentially depleted in comparison with heavier mass bodies. Since this reduction is mostly prevalent at shorter distances and we have an observational bias towards close-orbiting exoplanets; this decrease should be more pronounced. It is therefore possible that for smaller exoplanets; detection biases, more efficient ejections, and their predicted lack of fast type 1 migration could cancel out with their intrinsically larger populations which results in a perceived initial flat mass function. Despite our initial mass function differing from \citet{Owen2017}, we predict an almost identical distribution to \citet{Ginzburg2018} even though their mass-loss mechanism is considerably different to ours. Their model argues for a broken power-law which is flat from $\rm 0-5M_{\oplus}$ and then approaches zero for cores $\rm > 5M_{\oplus}$. We predict a slightly extended flat region but the overall shape is in strong agreement.

With regards to our final mass function, it grows rapidly for smaller masses which is compatible with the observations from \citet{Howard2010}. The rise in the abundance for bodies with $\rm \lesssim 3M_{\oplus}$ is due to a collection of planets without hydrogen envelops located in that region. Furthermore, the small maximum at $\rm \sim 8 M_{\oplus}$ is consistent with the observed mass peak of the \textit{Kepler} planets as shown by \citet{Marcy2014} and \citet{Malhotra2015}. However, even though the observed mass distribution is in agreement with our model, we are aware that the actual mass distribution of \textit{Kepler} planets has not been reliably determined due to the difficulty of obtaining the masses of small exoplanets.

In Figure~\ref{fig:collage} we show how the final radius and mass distribution would change depending on our initial mass function. Our simulations show that the best fit curve is a flat region from $\rm 3-8M_{\oplus}$ followed by a pareto distribution as shown in figure~\ref{fig:finalmassdistribution}. The truncated gaussian mass function shown in Figure~\ref{fig:collage} provides a possible alternative to our model especially because the observations by \citet{Fulton2017} have considerable uncertainties. However, we found that the major problem with a truncated gaussian function was the shape of the tail in relation to the position of the second radius peak. We were unable to get a fit where the location of the second peak lay at $\rm \sim 2.4R_{\oplus}$ whilst at the same time matching the shape of the tail of the distribution. We also tried to fit a truncated laplace distribution (not shown in the manuscript) to the initial masses but our results were similar (albeit slightly worse) to the truncated gaussian distribution. Finally, we show that without a deficiency in the occurance rate at larger masses (i.e. $\rm \gtrsim 8M_{\oplus}$), the bimodal distribution is strongly distorted and does not match the observations by \citet{Fulton2017}.

\subsection{Limitations of our Model}

When dealing with studies that are deeply theoretical such as this one it is crucial to consider the initial parameters and assumptions made since they can greatly affect the results. Due to the bimodal distribution of exoplanet radii being a multivariate phenomenon with a deeply stochastic nature it is not possible to test every parameter combination so assumptions were inevitable. One such assumption was that exoplanet orbital distances did not evolve and therefore remained constant. This is certainly false since many different effects can lead to large scale planetary migration such as (and not limited to):

\begin{enumerate}
	\item Gravitational scattering due to overdensities caused by other planets in the vicinity \citep[e.g.][]{Hansen2015}.
	\item In a binary system or when there are two planets with different inclinations, the Kozai mechanism can cause the planet in question to exchange eccentricity and inclination resulting in tidal friction and a subsequent shrinking of the orbital distance \citep[e.g.][]{Kozai1962,Nagasawa2008,Naoz2011}.
	\item Tidal migration. For example, a planet orbiting very close to its host star may induce a bulge which could lead to a loss in angular momentum if the star rotates faster than the planet's orbital period \citep[e.g.][]{Jackson2008,Penev2010,Penev2011}.
\end{enumerate}

Points 1 and 2 normally occur rapidly ($\ll 1$ Gyr) whilst 3 occurs on a longer timescale ($\rm \sim few$ Gyr). Other aspects we ignored include, but are not limited to: how for close-orbiting tidally locked super-Earths, the tidal forces, together with the orbital and rotational centrifugal forces, could partially confine a hydrogen-rich atmosphere on the nightside \citep{Modirrousta2020}; secondary atmospheres and how they affect the radii of exoplanets; meteorite impacts and how they can influence atmospheres \citep[e.g.][]{Miller2009,Lupu2014}; star mergers forming disks from which planets can form \citep[e.g.][]{Tutukov1991,Martin2011}; and planet-planet collisions \citep[e.g.][]{Ji2011,Chrenko2018}. These mechanisms are not only very hard to model, but most probably negligible and/or statistically insignificant.

\section{Conclusions}

After exposing $10^{6}$ synthetic exoplanets to their host star's XUV irradiation we show that the bimodal distribution observed by \citet{Fulton2017} can be reproduced. Our results indicate that for the radius gap to exist it is essential for there to be an initial paucity of exoplanets with masses $\rm \gtrsim 8M_{\oplus}$. Furthermore, our best fit results suggest that there is a flat distribution of exoplanets with masses $\rm 3M_{\oplus} - 8M_{\oplus} $ and a paucity for planets with masses $\rm \lesssim 3M_{\oplus}$. In other words, the initial distribution of exoplanet masses, has a a great influence on the final radius distribution. With regards to the properties of the radius distribution, we predict that the peak situated at $\rm \sim 1.3 R_{\oplus}$ consists mostly of rocky denuded bodies whilst the maximum at $\rm \sim 2.4 R_{\oplus}$ marks a region full of hydrogen-rich exoplanets with a few mega-Earths. There are some very rare exceptions such as metallic planets with small hydrogen atmospheres and denuded $\rm H_{2}O$ ice planets \citep{Zeng2018} which could lay in the first and second peak respectively. Finally, we believe that our predictions can be tested in the not too distant future due to rapid technological advances. For instance, the James Webb Space Telescope (JWST), Atmospheric Remote-sensing Infrared Exoplanet Large-survey (ARIEL), Thirty Meter Telescope (TMT), and Extremely Large Telescope (ELT) should become active in the next few years which would allow for far superior exoplanet observations. In the future, statistical methods on large data sets could be used in contrast to analysing individual cases like at the present. For instance, with Tier 1 of ARIEL we could test the exoplanets at each peak on whether or not they have an atmosphere; this would provide strong evidence for or against our results. Furthermore, if we can get more unbiased mass distribution measurements for super-Earths and sub-Neptunes through Earth-based observations, this would be a strong test for the validity of our model.


\section*{Acknowledgements}

We acknowledge the support of the ARIEL ASI-INAF agreement n.2018-22-HH.0. We thank P. Neague and the anonymous referee for their useful comments.

\bibliography{bibliography.bib}
\bibliographystyle{aasjournal}

\appendix

\section{Extra Information on the XUV Flux}

\begin{figure}[h]
	\centering
	\includegraphics[scale=0.6]{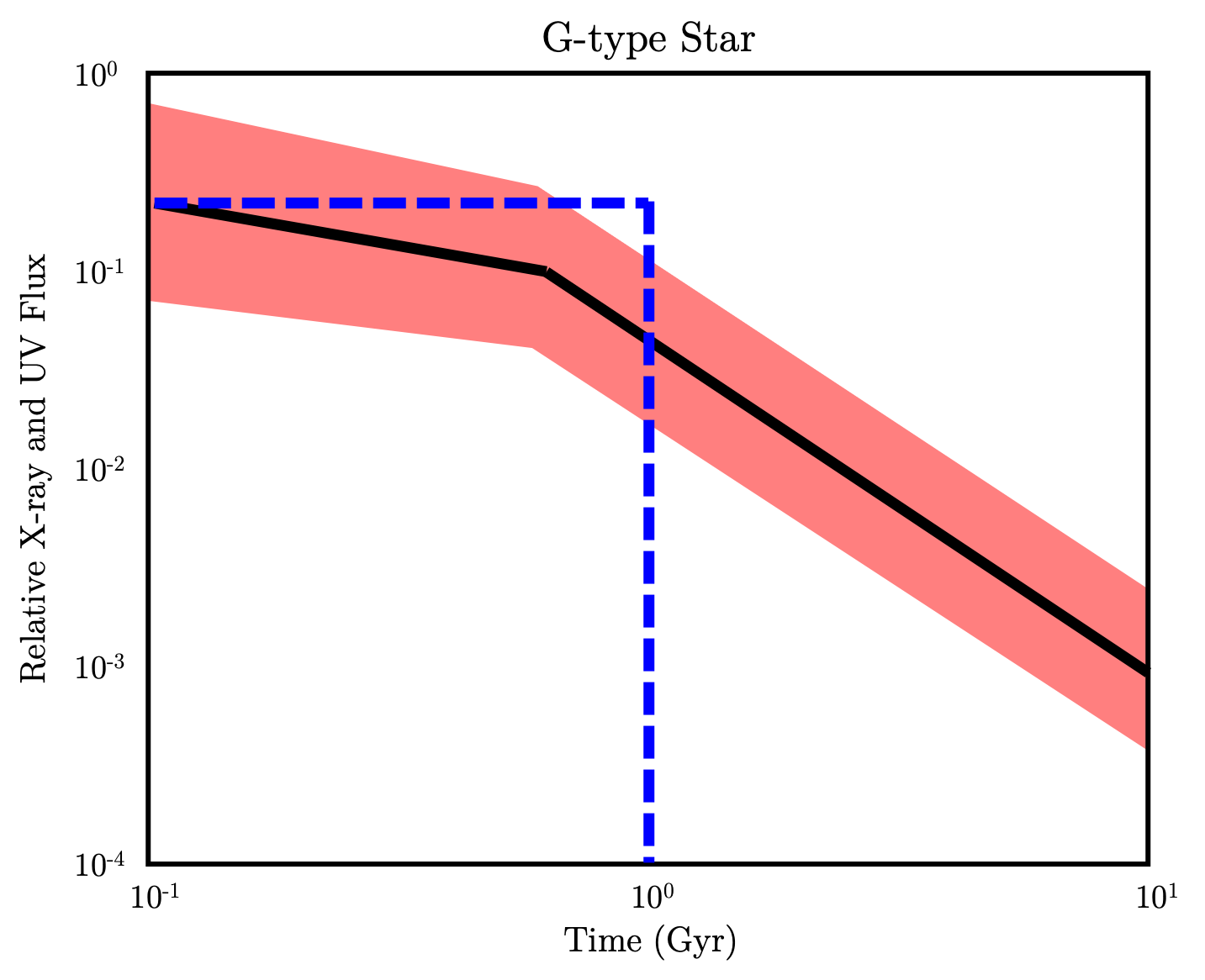}
	\caption{The relative XUV flux of G-type stars. The red shaded region marks the measurements from \citet{Penz2008(2)} and \citet{Sanz-Forcada2011}. The black line is the average measured XUV flux. The blue dashed line shows the step function we adopted in our model.}
	\label{fig:gtypexuv}
	\centering
\end{figure}

\begin{figure}[h]
	\centering
	\includegraphics[scale=0.6]{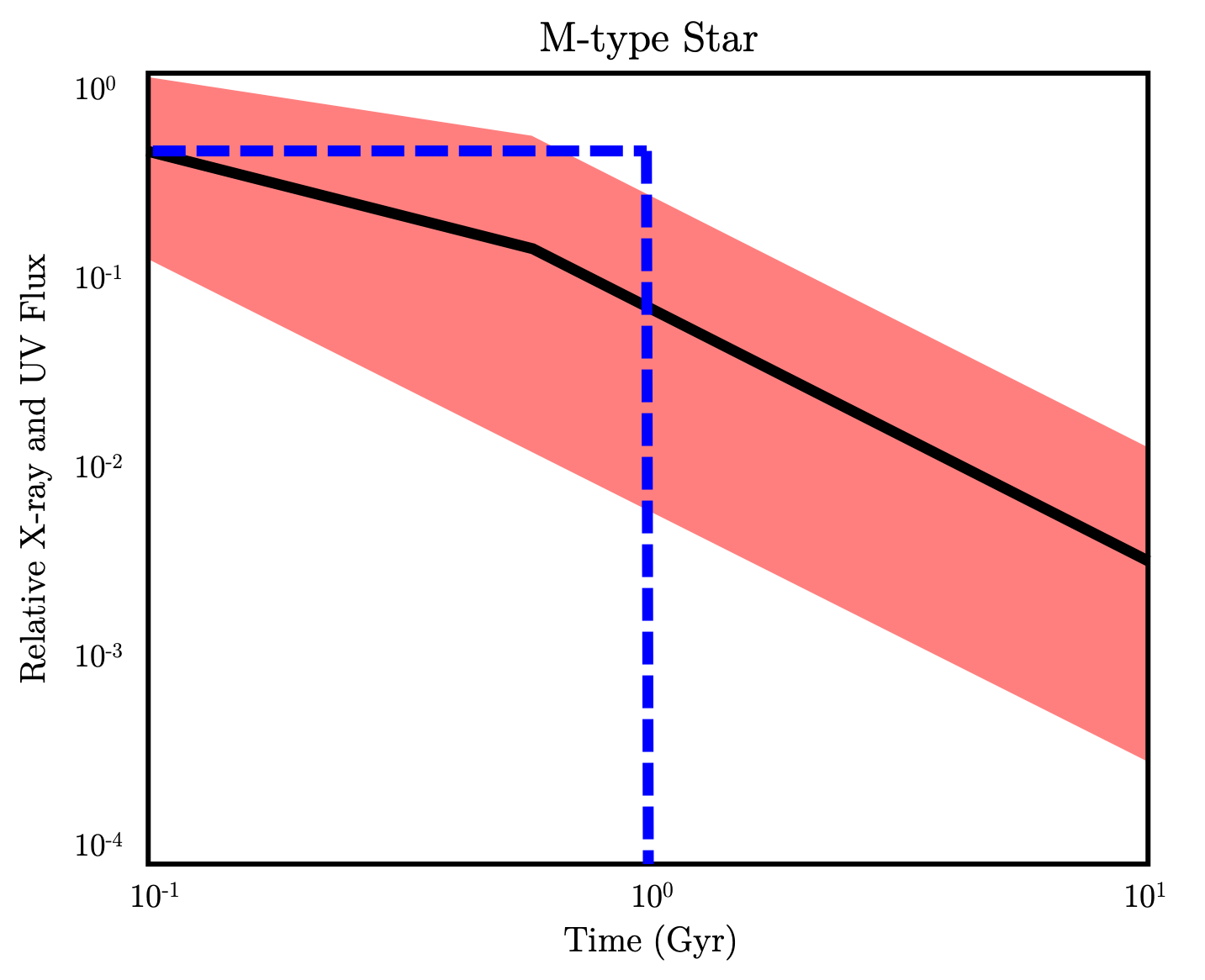}
	\caption{The relative XUV flux of M-type stars. The red shaded region marks the measurements from \citet{Penz2008(1)} and \citet{Sanz-Forcada2011}. The black line is the average measured XUV flux. The blue dashed line shows the step function we adopted in our model.}
	\label{fig:mtypexuv}
	\centering
\end{figure}

\end{document}